\documentclass{article}
\usepackage[utf8]{inputenc}
\usepackage{authblk}

\title {Why Do You Think It is a Black Hole?}
\author{Galina Weinstein\thanks{This work is supported by ERC advanced grant number 834735.}}
\affil{\normalsize The Department of Philosophy, University of Haifa, Haifa, the Interdisciplinary Center (IDC), Herzliya, Israel.} 

\date{March 2, 2021}

\begin{document}

\maketitle

\begin{abstract} 
This paper analyzes the experiment presented in 2019 by the Event Horizon Telescope (EHT) Collaboration that unveiled the first image of the supermassive black hole at the center of galaxy M87. The very first question asked by the EHT Collaboration was: What is the compact object at the center of galaxy M87? Does it have a horizon? Is it a Kerr black hole? In order to answer these questions, the EHT Collaboration first endorsed the working hypothesis that the central object is a black hole described by the Kerr metric, i.e. a spinning Kerr black hole as predicted by classical general relativity. They chose this hypothesis based on previous research and observations of the galaxy M87. After having adopted the Kerr black hole hypothesis, the EHT Collaboration proceeded to test it. They confronted this hypothesis with the data collected in the 2017 EHT experiment. They then compared the Kerr rotating black hole hypothesis with alternative explanations and finally found that their hypothesis was consistent with the data. In this paper, I describe the complex methods used to test the spinning Kerr black hole hypothesis. I conclude this paper with a discussion of the implications of the findings presented here with respect to Hawking radiation.

\end{abstract}

\section{Introduction}

In April 2019 an international collaboration of scientists (hereafter EHT Collaboration) unveiled an image that shows an apparently blurred asymmetric ring around a dark shadow. After
comparison with computer models, the picture was interpreted as showing a supermassive black hole at the center of the galaxy Messier 87 (hereafter M87). This interpretation fitted well with theory and simulation, according to which the M87 black hole is one of the two largest supermassive black holes on the sky along with the Sagittarius A (hereafter Sgr A) black hole (at the center of the Milky Way, near the border of the constellations Sagittarius and Scorpius).

Shepherd Doeleman, the founding leader of the Event Horizon Telescope (hereafter EHT) project, has recently explained that the EHT is “a global Project that came together with
a specific focused scientific objective in mind, which was to make the first image of a black hole". 
Doeleman further remarked: “you can see here the horizontal bands showing the locations of those stars that they [Eddington and his team] used to determine this […] and now here we are just over hundred years later, thinking of doing the same kind of experiment but around the black hole” \cite{Doeleman}.

The Arthur Eddington solar eclipse expeditions seem to have inspired the EHT Collaboration to do their experiment. The 1919 eclipse expeditions confirmed Einstein's 1915 prediction of the deflection of light near the Sun. An astrographic telescope was used for the photography of the field of stars surrounding the Sun in the 1919 total eclipse. Eddington’s famous total solar eclipse photograph shows the eclipsed Sun (a round large black “shadow”) surrounded by a few displaced stars. The locations of the displaced stars are indicated by a few white horizontal bands. Although the supermassive black hole does not behave like an eclipsed Sun, light is nevertheless deflected by both objects.

The very first question asked by the EHT Collaboration was: What is the compact object at the center of galaxy M87? Does it have a horizon? Is it a Kerr black hole? In order to answer these questions, the EHT Collaboration first reconstructed images from data obtained in the experiment, as discussed in Section \ref{Section 2}. They then endorsed the working hypothesis that the central object is a black hole described by the Kerr metric, i.e. a spinning Kerr black hole as predicted by classical general relativity, as discussed in Section \ref{Section 3}. The EHT Collaboration chose this hypothesis based on previous research and observations of the galaxy M87. After having adopted the Kerr metric, they proceeded to test it. They confronted this hypothesis with the data collected in the 2017 EHT experiment. In Sections \ref{Section 3}-\ref{Section 6}, I describe the complex methods used to test the Kerr black hole hypothesis. The EHT Collaboration then compared the Kerr rotating black hole hypothesis with alternative explanations and finally found that their hypothesis was consistent with the data, as shown in Section \ref{Section 6}. 

I analyze the black hole image and the EHT experiment on the basis of the six letters published by the EHT Collaboration in \emph{The Astrophysical Journal Letters} in April 2019, which contain a complete exposition of the experiment \cite{EHTCa}-\cite{EHTCf}. 

In section \ref{Section 8}, I discuss a paper (letter) submitted at about the same time as the six EHT \cite{EHTCa}-\cite{EHTCf} letters. The paper discusses the GW150914 gravitational wave event. The authors also adopt the Kerr black hole hypothesis and test it. In both cases – the EHT experiment and the gravitational wave experiment – the Kerr black hole hypothesis is selected as the best explanation. I discuss the philosophical implications of this selection in Section \ref{Section 7}. I end this paper with Section \ref{Section 9}, discussing the philosophical significance of the findings presented here with respect to Hawking radiation.

\section{Evidence and interpretation of the evidence} \label{Section 2}

In 2000, Heino Falcke et al. performed simulations of an optically thin emitting accretion disk surrounding a maximally spinning supermassive Kerr black hole with spin $a_*=0.998$ and mass $M_{BH}=2.6 \times 10^6 M\odot$ [solar masses ($M\odot$), $a_*$ is defined by equation (\ref{equation1}) in Section \ref{Section 3}]. They presented ray traced images and images seen for a wavelength of 1.3 mm and wrote that “a marked deficit of the observed intensity inside the apparent boundary” is produced “which we refer to as the ‘shadow’ of the black hole”. They further noted that to simulate the images we need a Very Long Baseline Interferometry (hereafter VLBI) array of radio telescopes located $\approx 8,000 km$ apart \cite{Falcke}, p. L14. 

Event-horizon observations of the core of the galaxy M87 were conducted on the two consecutive nights of April 5/6, 2017 and again four nights later on the two consecutive nights of April 10/11, 2017. The observations were made at a wavelength of 1.3 mm and from a VLBI array of radio telescopes located $\approx 10,000 km$ apart. Scan durations of M87 varied between three and seven minutes and the quasar 3C 279 was also scanned as a calibrator source. The compact object at the center of M87 was observed with eight telescopes at six geographic sites and it was the first time that the small array called Atacama Large Millimeter/submillimeter Array (hereafter ALMA) has been included in the VLBI array. This whole array is referred to as the EHT. 

After Correlation, the visibility $V_{ij}$ data were calibrated by three calibration pipelines \texttt{EHT-HOPS}, \texttt{rPICARD} and \texttt{AIPS} (Python-based scripts using modules from the EHT analysis toolkit, the \texttt{eat} library).\footnote{The radio signals collected by each telescope are digitized and labeled with timestamps created by atomic clocks at each site and then recorded to hard drives. The data are read from each hard drive to a correlator and then processed by the correlator’s software, which matches up and compares the data streams from every possible pairing of the EHT's eight telescopes. After correlation, the complex visibility data $V_{ij}$ are obtained. $V_{ij}$ is the output of the correlator and the fundamental data product which gives information of both the amplitude and the phase of the fringes of the signal on a baseline (distance between two telescopes/stations) $i$ and $j$.} Phase errors (fringe-fitting) and atmospheric (turbulence and variations) errors were corrected and fixed – and the amount of data was consequently reduced so that noisy data was removed. All three pipelines produced data – visibility amplitudes as a function of baseline – with a slight anisotropy. One can see peaks in the observed visibility amplitudes that appear across all four observed days. These peaks have two minima, nulls, on either side and at the location of the minima the visibility amplitudes of the data are very low. The first of the nulls occurs at $\sim 3.4 G \lambda$ (giga-lambda) and the second is observed at $\sim 8.3 G \lambda$. The high peak between these two nulls is at $\sim 6 G \lambda$. A single pipeline output, the \texttt{EHT-HOPS}, was designated as the primary data set of the engineering data release \cite{EHTCb}, p. 14; \cite{EHTCc}, p. 2, 5-8, pp. 23-24.

Residual station-based amplitude and phase calibration errors nevertheless persist after performing calibration steps and overwhelmingly dominate the remaining thermal random Gaussian noise $\epsilon$. It is therefore possible to construct data products that are insensitive to these systematic errors. Such quantities are called closure quantities.\footnote{At high frequencies, visibility amplitude and phase calibration are made extremely difficult. Amplitude gain ($g_i$ and $g_j$) and phase ($\Phi_i$ and $\Phi_j$) errors corrupt the measured visibility: $V_{ij}=g_i g_j {\mathcal{V}}_{ij} e^{i(\Phi_i-\Phi_j)}+ \epsilon$, where ${\mathcal{V}}_{ij}$ stands for the unknown model visibility (the ideal measured visibility with no atmospheric corruption). Radio waves from M87 reach station $i$ before reaching station $j$. Closure phase is the product of three visibilities (from three stations). It is formed from baseline visibilities on a closed triangle $ijk$. Consider three telescopes $ijk$. The visibility we measure on baseline $ij$ is: ($u$,$v$)$ij$ plus phase error at the station $(i-j)$. The measured visibility on baseline $jk$ is: ($u$,$v$)$jk$ plus phase error $(j-k)$ and the one measured on baseline $ki$ is: ($u$,$v$)$ki$ plus phase error $(k-i)$. Now the phase errors ($i-j$) $+$ ($j-k$) $+$ $(k-i)$ are all canceled when the phase visibilities are added up, leaving a quantity: $(u, v){ij}$ $+$ $(u, v){jk}$ $+$ $(u, v){ki}$ such that the baselines “close”. The logarithmic closure amplitude is constructed from combinations of visibilities measured on four stations and is insensitive to variations in the amplitude gains \cite{EHTCd}, pp. 2-4.}

In June 2018 the EHT Collaboration split up into four teams in different regions of the world and each team chose a different imaging method and worked in isolation for seven weeks on the data. The 2017 EHT April 11 data was selected.\footnote{The reason for selecting the 2017 EHT April 11 data set was that, on that observing day, the EHT array of radio antennas covered the largest area for M87 and the amplitude calibration among stations was the most stable.} The teams neither talked nor crossed photos and that enabled them to avoid getting into collective bias influencing their final images. Each one of the four teams relied upon the judgment of its members to select a different imaging method to convert the data into images: teams 1 and 2 used the Regularized Maximum Likelihood (hereafter RML) method, while Teams 3 and 4 used a version of the CLEAN algorithm \cite{EHTCd}, p. 9.\footnote{The two imaging techniques used in generating the 2017 EHT black hole images are based on the traditional deconvolution CLEAN algorithm and the RML method. Imaging algorithms are broadly categorized into two methodologies: inverse modeling (CLEAN) and forward modeling (RML). 1) CLEAN: The sky is only sparsely sampled by the VLBI array. The original CLEAN algorithm begins with a Fourier transform of the sampled visibilities (the dirty image). The algorithm then proceeds to find the points of highest intensity in the dirty image. The process is iterative and iterations are continued until a desired noise level (a stopping criterion) is achieved in the dirty image. Finally, CLEAN takes the accumulated point source model and convolves the image with a restoring beam, a clean beam, Gaussian beam of full width at half maximum (FWHM), instead of the dirty beam. The algorithm adds the residuals of the dirty image to this image to form the CLEAN image. 2) RML is based on maximum likelihood estimation (MLE): finding the best fit image that minimizes a function that is a sum of a chi-squared $\chi^2$ function corresponding to data and another function called regularizer. Regularization may include smoothness (requiring that the image be smooth) or sparsity) \cite{EHTCa}, p. 4; \cite{EHTCd}, p. 4.}

In July, 2018, all four teams came together and revealed for the first time the four images they had produced. Although not precisely identical, when comparing images, all four teams
obtained an image of a crescent with a central dark area surrounded by an asymmetric ring. The initial blind imaging stage indicated that the image was dominated by an angular ring diameter (an angular crescent diameter) of $\theta_d \sim 40 \mu as$ \cite{EHTCd}, p. 9.

The next stage of the experiment required the objective  evaluation of the fidelity of the images reconstructed in the first stage. That is, the second stage involved selecting imaging parameters that are independent of expert judgment.

Three imaging Python pipelines, \texttt{DIFMAP} (Difference Mapping), \texttt{SMILI} (Sparse Modeling Imaging Library) and \texttt{eht-imaging}, have been designed. The \texttt{SMILI} and \texttt{eht-imaging} pipelines are based on RML and \texttt{DIFMAP} is a CLEAN Python script.\footnote{\texttt{DIFMAP} is a CLEAN Python script implemented in DIFMAP, a software package originally written by Martin Shepherd in the 1990s. After loops of cleaning and self-calibration, the \texttt{DIFMAP} script generates final cleaned images. The \texttt{SMILI} and \texttt{eht-imaging} pipelines have been developed by the EHT Collaboration. The \texttt{SMILI} Python-interfaced library is based on sparse sampling, reconstruction of images from sparse data. NumPy and SciPy power the main tasks of \texttt{eht-imaging} and the main imaging methods are powered by the routine \texttt{scipy.optimize.minimize}. Other libraries used in creating the 2017 EHT image are: Astropy, Pandas, and Matplotlib by which the black hole image is visualized and processed.} The pipelines employ an iterative imaging loop, alternately imaging and self-calibrating the data (visibility amplitudes and closure quantities)\cite{EHTCd}, p. 12.\footnote{The \texttt{eht-imaging} pipeline handles self-calibration by minimizing:

\vspace{1mm} %1mm vertical space
\noindent $\chi^2= \sum_{i<j}\frac{\left(\mid V\mid_{ij}-g_ig_j\mid {\mathcal{V}}\mid_{ij}\right)^2}{\sigma_{ij}^2}$ 
\vspace{1mm} %1mm vertical space
\noindent for each baseline in the EHT array. 
$\mid V\mid_{ij}$ is the measured visibility amplitude, $\mid{\mathcal{V}}\mid_{ij}$ stands for the model visibility amplitude and $\sigma_{ij}^2$ is the uncertainty on $\mid V\mid_{ij}$. While in progress, the imaging + self-calibrating pipelines gradually remove $g_i$ and $g_j$ by going through a loop of multiple iterations.}

Four training geometric models of sources with different compact structure (a ring, a crescent, a disk and two blobs) were considered. The models were chosen such that their visibility amplitude data on EHT baselines matched salient features of the observations, in particular the first visibility null at at $\sim 3.4 G \lambda$ and the second at $\sim 8.3 G \lambda$. You might rightly think that a disk can definitely not fit the 2017 EHT data. But all four models produce images of objects having an apparent diameter of $\approx 50\mu as$. Further, all of the geometric compact objects have a total flux density of $0.6$ Jy and also a $0.6$ Jy jet that is resolved by EHT baselines. Synthetic data were generated from the model images of the four different geometric models using two independent Python imaging pipelines, \texttt{MeqSilhouette} and \texttt{eht-imaging}. While \texttt{MeqSilhouette} was used for comparison, \texttt{eht-imaging}’s synthetic data were used for image-reconstruction.

Each one of the three pipelines, \texttt{DIFMAP}, \texttt{eht-imaging} and \texttt{SMILI}, has fixed settings (such as the pixel size) and also parameters taken as arguments.\footnote{\texttt{DIFMAP} has five parameters taken as arguments: the total flux density; the stop condition (imaging stops when the desired flux density is reached or when the decrease in the residual image rms noise dips below
a given threshold); a relative correction factor for ALMA in the self-calibration process, the diameter of the cleaning region ($40$ to $100\mu as$) and a parameter controlling the visibility errors. \texttt{eht-imaging} and \texttt{SMILI} both have seven parameters as arguments: the total flux density; the fractional systematic error on
the measured visibilities, the FWHM of the circular Gaussian initial image, and four regularization terms.} 
Those imaging parameter combinations for the three different imaging pipelines were derived by conducting parameter space surveys on synthetic data. To determine a single combination of the best performing hyper parameters for each pipeline, called fiducial parameters, many similar-looking images were reconstructed from the synthetic data (produced by \texttt{eht-imaging}). 

The images were reconstructed using $50,000$ possible parameter combinations, which also included values that were expected to produce poor reconstructions. The images were then compared with images that were generated from the 2017 EHT data. Cross-correlation between synthetic images and 2017 EHT images was computed and values were given to perfect correlation between images, no correlation, and perfect anticorrelation. The fidelity of the synthetic images was quantified by the FWHM Gaussian beam (blurring) between the synthetic images and the 2017 EHT images. A value of the blurring, the FWHM Gaussian beam, for each combination of imaging parameters was then computed, and the combination of parameters was ranked according to this value. The top-ranked combination of parameters was then denoted as the “fiducial parameters” of that method. The Top Set of imaging parameter combinations for each pipeline was selected. 

The hyper-parameter selection procedure was subsequently tested. The purpose of the test was to see whether or not the chosen parameters perform well across geometric models, including geometric shapes that do not resemble the crescent observed in the 2017 EHT images. The disk model was excluded from the training set. All three imaging pipelines nonetheless produced from the synthetic data an image with a disk shape that did not resemble any of the other models. The fiducial parameters that best produced images of the disk shape were those that described the crescent, the ring, and the double objects models. By repeating the same process with each of the four geometric models, the set of fiducial imaging hyper-parameters was cross-validated. The chosen fiducial parameters were able to distinguish between the four geometric shapes and perform well on average in reconstructing images of the four shapes, of what was there.

The three pipelines were now fed with the April 11, 2017 EHT data set and produced the same fundamental structure: an asymmetric ring around a dark shadow. Fiducial images of a yellow orange asymmetric bright ring around a central dark shadow (a crescent) were reconstructed from the April 11, 2017 EHT data set.\footnote{A Python module \texttt{ehtplot} with a color submodule has been created, a perceptually uniform colourmap that is used in the reconstruction and simulations of images. The colors of the images represent the brightness temperature which does not necessarily correspond to any physical temperature of the radio emitting plasma in the jets and accretion disk \cite{EHTCd}, pp. 8-9.} Reconstructed images of data sets from observations made on April 5, 6, and 10 were subsequently incorporated, while preventing the poorest reconstructions of
2017 EHT April 10 from dominating the outcome. The best fiducial 2017 EHT images, for each of the four observed days, from each of the three imaging pipelines, were subsequently selected, the averages of which were taken and restored to an equivalent resolution of the EHT array. The images produced from the three pipelines are broadly consistent across all four observing days and show a central shadow and a prominent asymmetric ring having an enhanced brightness toward the south. The EHT Collaboration formed a single image, which they called the \emph{consensus image}, from the average of each day. This image is the famous black hole image of M87 \cite{EHTCa}, p. 5; \cite{EHTCd}, pp. 7-21, p. 22, pp. 32-33, pp. 37-41; \cite{EHTCf}, p. 46; \cite{Weinstein}, p. pp. 5-6. 

Having reconstructed images by \texttt{DIFMAP}, \texttt{SMILI} and \texttt{eht-imaging}, validation tests were performed to assess their reliability. Recall that in April 2017, observations of the calibrator 3C 279 were performed with the EHT. Images of 3C 279 were reconstructed and self-calibrated by \texttt{DIFMAP}, \texttt{SMILI} and \texttt{eht-imaging} and then compared with the 2017 EHT M87 images. It was shown that station gains in both cases were broadly consistent. It was then concluded that consistency between the two sources M87 and 3C279 and among the three different pipelines provides confidence that the gain corrections are not imaging artifacts or missing structures in the 2017 EHT images \cite{EHTCd}, pp. 22-24.

Finally, reconstructed images were generated using \texttt{SMILI}, \texttt{eht-imaging} and \texttt{DIFMAP}, the purpose of which was to extract $\theta_d$. $\theta_d$ was extracted in the image domain of the fiducial images, for each of the four observed days, from each imaging pipeline (\texttt{SMILI}, \texttt{eht-imaging} and \texttt{DIFMAP}). The images were generated with the fiducial hyper-parameters. Unlike the \texttt{SMILI} and \texttt{eht-imaging} images, the ones obtained by \texttt{DIFMAP} were restored with a FWHM Gaussian $20\mu as$ beam. The measured image domain ring diameters $\theta_d$ were nonetheless consistent among all imaging pipelines. Across all days, the \texttt{DIFMAP} images recovered an average value of $\theta_d \approx 40 \pm 3\mu as$ when restored with a $20\mu as$. The \texttt{SMILI} and \texttt{eht-imaging} images recovered an average value of $\theta_d \approx 41 \pm 2\mu as$.

It is written at the end of the fourth letter: “These first images from the EHT achieve the highest angular resolution in the history of ground-based VLBI” \cite{EHTCd}, pp. 29-31. This is certainly true. The 2017 EHT data were subsequently interpreted. 

\section{Comparing the merits of two rival hypotheses} \label{Section 3}

In the fifth letter, the EHT Collaboration is endorsing the hypothesis that at the center of M87 there is a supermassive Kerr rotating black hole \cite{EHTCe}, p. 1. The Kerr solution of the Einstein field equations describes the spacetime geometry around astrophysical rotating black holes. An alternative hypothesis for what could be the compact object at the center of M87 had been checked and discarded before the 2017 EHT experiment was performed. This alternative is the hypothesis that the compact object at the center of M87 has no horizon but has a surface (such compact object could be very dense neutron stars and supermassive stars).

In 2015 a team of scientists belonging to the EHT Collaboration made an experiment the goal of which was to compare the merits of the two rival hypotheses: a compact object with a surface and a black hole with a horizon; and then choose which hypothesis can best explain the evidence. At the time in which the experiment was conducted, the EHT consisted of only three stations (radio telescopes). Were a compact object with a surface present at the center of M87 instead of a black hole with a horizon, said the team, its photosphere would be heated by the constant deposition of kinetic energy from the accretion gas. In the absence of an event horizon, the photosphere would therefore radiate, resulting in an additional component in the measured spectrum. The surface luminosity is proportional to the average mass accretion rate on the surface. The implication is that if the object at the center of M87 has a horizon, the kinetic energy of the accreting gas must advect past an event horizon, beyond which it is no longer visible to distant observers. In simple terms, a compact object with a surface would appear brighter than a black hole with a horizon. It was found that the existence of an observable photosphere in the core of M87 was ruled out. “The implication is that the kinetic energy of the gas is advected past an event horizon, beyond which it is no longer visible to distant observers. In other words, M87* must have an event horizon” \cite{Broderick2}, pp. 7-8, and the black hole hypothesis was selected.

In the fifth letter, it is written: “In this Letter we adopt the working hypothesis that the central object is a black hole described by the Kerr metric, with mass $M$ and dimensionless spin $a_*$”. After having adopted the Kerr metric and the no-hair theorem, the EHT Collaboration proceeds to test the Kerr metric by fitting GRMHD models to 2017 EHT data.\footnote{The Kerr metric was proved to be the unique stationary, asymptotically flat, vacuum solution with an event horizon. It was shown that the Kerr-Newman solution involves just three free parameters: mass $M_{BH}$, angular momentum $J$ (spin $a_*$) and charge $Q$. This is the no-hair theorem which states that non-charged black holes are uniquely characterized by their mass $M_{BH}$ and angular momentum $J$ and are described by the Kerr metric. The black hole is governed by general relativity (GR). Magnetohydrodynamics (MHD) is the framework that governs the dynamics of the accretion flow and jets around the black hole. The plasma is treated as a fluid and numerical methods to integrate the GRMHD equations are searched for.}

A GRMHD Simulation Library has been generated from several different codes.\footnote{1) The Black Hole Accretion Code (BHAC) performs magnetohydrodynamical simulations of an accretion flow onto a black hole. The ions (protons) and the electrons in the accretion disk plasma travel quite a long distance along magnetic field lines before being scattered. This and other factors complicate the calculations and require more computational power. 2) A system called H-AMR (Hierarchical Adaptive Mesh Refinement) is used, which accelerates GRMHD calculations by implementing the AMR strategy: a method that reduces the number of computations. 3) Two additional codes, a 3D version of High Accuracy Relativistic Magnetohydrodynamics (iharm3D) and 4) KORAL [Kod radiacyjny L (in Polish)], solve the GRMHD conservation laws by a shock-capturing method \cite{Porth}, pp.  6–8, p. 31.} Forty-three high-resolution, three-dimensional Standard and Normal Evolution (SANE) and Magnetically Arrested Disk (MAD) simulations, covering well the physical properties of magnetized accretion flows onto Kerr black holes, were performed by varying two dimensionless parameters:
\vspace{1mm} %1mm vertical space

1) The spin $a_*$:

\begin{equation} \label{equation1}
a_*\equiv \frac{Jc}{GM_{BH}^2}, -1<a_*<1.
\end{equation}

\vspace{1mm} %1mm vertical space
\noindent where $G$ is the universal gravitational constant, $c$ is the velocity of light. 
\vspace{1mm} %1mm vertical space

\noindent If $G = c = 1$, then: $a_*\approx \frac{J}{M_{BH}^2}$.

\vspace{3mm} %3mm vertical space

2) The magnetic flux $\phi$:\footnote{The SANE and MAD models are an approximation of the complex non-linear general relativistic dynamical system of a magnetized accretion disk flow orbiting a rotating supermassive black hole. Models with $\phi \approx 1$ in equation (\ref{equation2}) in the text are conventionally referred to as SANE where accretion is largely unaffected by the black hole magnetic field. If the magnetic flux reaches $\phi \approx 15$, the large-scale magnetic field accumulated by the accretion disk stops the accretion flow. Models with $\phi \approx 15$ are conventionally referred to as MAD. At $\phi > \approx 15$, numerical simulations show that the accumulated magnetic flux erupts, pushes aside the accretion flow, and escapes \cite{EHTCe}, p. 4.} 

\begin{equation} \label{equation2}
\phi=\frac{\Phi_{BH}}{\sqrt{\dot{M} {R_g}^2}c}, 1 \leq \phi < 15.
\end{equation}
   
\noindent where $\Phi_{BH}$ is the magnetic flux, $\dot{M}$ is the mass accretion rate (the rate at which mass is accreted onto a black hole) and $R_g$ is the gravitational radius defined by equation (\ref{equation4}) below.

The Simulation Library contains SANE models with $a_*= -0.94$, $-0.5$, $0$, $0.5$, $0.75$, $0.88$, $0.94$, $0.97$, and $0.98$, and MAD models with $a_*= -0.94$, $-0.5$, $0$, $0.5$, $0.75$, and $0.94$.

The location of the surface of the event horizon of the Kerr black hole is:

\begin{equation} \label{equation3}
r_+=R_g \left(1+ \sqrt{1-a_*^2} \right).
\end{equation}

\noindent where $R_g$ is given by:

\begin{equation} \label{equation4}
R_g= \frac{G {M_{BH}}}{c^2}.
\end{equation}

\vspace{1mm} %1mm vertical space
In order to produce images from the simulations, the following parameters were fixed in the GRMHD simulations:
\vspace{2mm} %2mm vertical space

1) \textbf{PA}. The position angle of the forward radio jet measured east of north. Based on prior knowledge from observations, the chosen value was: $PA \approx 288^0$.

2) \textbf{i}. The observer inclination, the orientation of the observer through $PA$. Based on prior knowledge from observations, images were generated at $i = 12^0$, $17^0$, $22^0$, $158^0$, $163^0$, and $168^0$ and a few at $i = 148^0$. Basically, images were generated at two major inclinations $17^0$ and $163^0$.

3) $\mathbf{\theta_g}$. The image scale:

\begin{equation} \label{equation5}
\theta_g= \frac{G {M_{BH}}}{dc^2}.
\end{equation}

\vspace{1mm} %1mm vertical space
\noindent If $G=c=1$, then $\theta_g\approx \frac{M_{BH}}{d}$, 
\vspace{1mm} %1mm vertical space
where $d$ is the distance to M87: $d=16.8 \pm 0.8$ Mpc (Megaparsec) and $\theta_g$ represents the angular gravitational radius. 

For $M_{BH}$ “we use the most likely value from the stellar absorption-line work, $M_{BH} \approx 6.2 \times 10^9 M\odot$ (\cite{Gebhardt}), see explanation in Section 5].

\noindent Inserting the above value of $M_{BH}$ into equation (\ref{equation9}) from Section \ref{equation5}:
\vspace{3mm} %3mm vertical space

\noindent $\theta_d= \frac{9.79 {M_{BH}}}{d} \approx 35.5 \mu as$
\vspace{3mm} is calculated. %3mm vertical space

4) $\mathbf{F_\nu}$. The total compact flux density measured in Jy (Jansky). The average flux density of $1.3 mm$ ($230$GHz) emission is $0.6$ Jy; 

5) $\mathbf{R_{high}}$. The temperature ratio of electrons to protons.\footnote{The plasma in the accretion disk is an accretion flow which is composed of ions and electrons. Both species have the same temperature in the funnel (the strongly magnetized regions of the accretion flow), but have a substantially different temperature in the middle of the accretion disk (the weakly magnetized regions). Assuming the gas is composed of non relativistic ions with temperature $T_i$ and relativistic electrons with temperature $T_e$, the ratio of the temperatures of the two species can be imposed in terms of a single parameter $R_{high}$. Relativistic electrons emit radio photons at $1.3 mm$ wavelength observed by the EHT known as synchrotron radiation. If the synchrotron radiation is emitted from weakly magnetized regions: $T_i \approx T_e$. If the emission comes from the funnel: $\frac{T_i}{T_e} \approx R_{high}$ \cite{EHTCe}, pp. 4-5, p. 12.} Images were generated at: ${R_{high}}=1$, $10$, $20$, $40$, $80$, and $160$. 

\vspace{2mm} %2mm vertical space
After fixing the values of these parameters ($\frac{M_{BH}}{d}$, $i$, $PA$, $F_\nu$, $R_{high}$) snapshots were drawn from the time evolution of the simulation at a cadence of every: 
\vspace{1mm} %1mm vertical space
$10-50 \frac{R_g}{c}$.

\vspace{1mm} %1mm vertical space
From the SANE and MAD simulations, more than 60,000 synthetic $1.3mm$ snapshot images were produced by three general relativistic ray-tracing (GRRT) codes, \texttt{ipole}, \texttt{RAPTOR} and \texttt{BHOSS}. The snapshot images created by GRRT codes show the black hole with a variety of accretion flows and jets and depict a yellow-orange asymmetric bright ring around a central dark shadow (a crescent). 

Each snapshot image generates a single snapshot model (SSM) defined by three parameters: $F_\nu, \theta_g, PA$. The snapshot images were compared with the 2017 EHT April 6 data set.\footnote{Model-fitting requires a large number of scans. The number of scans obtained of M87’s compact object each night ranged from seven on April 10 to twenty-five on April 6.}

\section{Models of accretion around a Kerr black hole} \label{Section 4}

Comparison of models to data was performed by computing the distance $\chi_\nu^2$ between the data and the snapshot image.\footnote{The reduced chi-squared statistic $\chi_\nu^2$. $\chi_\nu^2 \equiv \frac{\chi}{\nu} \equiv \frac{\chi}{N-M}$, where $\nu$ represents the degrees of freedom, $N$ is the number of data values and $M$ is the number of free parameters.} In the course of computing $\chi_\nu^2$, the three parameters $F_\nu, \theta_g, PA$ and the gains $g_i$ at each VLBI station are varied “in order to give each image every opportunity to fit the data” \cite{EHTCe}, p. 8. 

Variations in the three parameters $F_\nu, \theta_g, PA$ approximately correspond to variations in the accretion rate, black hole mass, and orientation of the black hole spin, respectively. In fitting GRMHD snapshot images to data, the image is stretched by adjusting $\theta_g \left( \frac{M_{BH}}{D} \right)$, re-scaled by varying $F_\nu$ and rotated by changing $PA$. Varying the parameters $a_*$, $\phi$ and $R_{high}$ can change the width and asymmetry of the photon ring and introduce additional structures exterior and interior to the photon ring \cite{EHTCe}, pp. 7-8.

Fitting models to data was performed by two different methods:
\texttt{GENA} (a genetic algorithm)\footnote{\texttt{GENA} fits models to data by finding the model parameters which minimize a $\chi_\nu^2$ statistic. It implements the differential evolution (DE) algorithm, an optimization algorithm inspired by natural evolution theory. The major steps of DE are the following: A random population of model parameters (individuals, parents) is created. Fitness of parameters with the data (visibility amplitudes and closure phases) is computed. Fitness is inversely proportional to $\chi_\nu^2$. The fitness of $F_\nu, \theta_g, PA$ with the data is computed such that maximizing fitness of parameters to data minimizes the value of $\chi_\nu^2$. In genetic algorithms, based on their fitness, best-fit parameters (parents) are improved (selected) by iterations in which mutation is used. \texttt{GENA}, however, utilizes a Nondominated Sorting GA \texttt{(NSGA)}, the \texttt{NSGA-II}, to explore the parameter space. \texttt{NSGA} implements an elitist selection process for multi-objective optimization. A population of parents is generated randomly. Offsprings are compared with parents. If the offsprings do not dominate their parents, they are sorted, moved to the next generation and the parents are replaced. If the offsprings dominate their parents, they are not elite offsprings and are rejected. The process of selecting the non-dominated offsprings continues until the initial population of parents is replaced. \texttt{GENA} further uses the procedure of network amplitude calibration of the residual station gains $g_i$, $g_j$ performed by the \texttt{eht-imaging} pipeline.}. And \texttt{THEMIS}, a Bayesian parameter estimation and model comparison framework, which has been specifically developed for the EHT experiment.\footnote{\texttt{THEMIS} uses the (differential evolution) DE-MCMC (Markov chain Monte Carlo) sampler in which many chains are run in parallel. An MCMC sampler is a random walk method through parameter space for performing Bayesian inference. It randomly samples the posterior probability distribution of parameters and generates a sequence of random samples of parameters that fits the data. The parameter is dependent upon the previous one in the chain. DE-MCMC is especially efficient in sampling from models with highly correlated parameters. \texttt{THEMIS} implements a parallel-tempering algorithm for DE-MCMC. Parallel tempering is a method based on an analogy with statistical physics, called thermodynamic integration, to calculate the Bayesian evidence. Many MCMC chains are run in parallel on tempered versions of the original likelihood function at different temperatures. Within \texttt{THEMIS} at each MCMC step, the station gains $g_i$, $g_j$ are addressed by marginalizing over the nuisance parameters. Gain amplitudes $g_i$ represent approximately between $40$ and $143$ additional nuisance parameters per data set. The gain parameters are subsumed into the likelihood and incorporated as model parameters. Assuming
Gaussian priors, the log-likelihood: 
\vspace{1mm} %1mm vertical space

$\mathcal{L}= -\sum_{i<j}\frac{\left(\mid V\mid_{ij}-g_ig_j\mid {\mathcal{V}}\mid_{ij}\right)^2}{2\sigma_{ij}^2}$ 

\vspace{1mm} %1mm vertical space
\noindent is maximized and a marginalization over all $g_i$ is performed \cite{Broderick1}, pp. 14-15, p. 35; \cite{EHTCf}, pp. 5-6, p. 25.} \texttt{THEMIS} uses a Differential Evolution Markov Chain Monte Carlo (DE-MCMC) sampler to produce Bayesian posterior estimates for the three
parameters $F_\nu, \theta_g, PA$ whilst \texttt{GENA} uses a DE algorithm for producing best fit estimates for the three parameters \cite{Broderick1},  p. 4; pp. 15-16; \cite{EHTCa}, pp. 3-6, p. 10, p. 36; \cite{EHTCd}, pp. 8-9, p. 14, p. 28; \cite{EHTCe}, p. 3, pp. 7-8; \cite{EHTCf}, p. 6, p. 15. 

Computing the distance $\chi_\nu^2$ between the data and the snapshot images is a difficult challenge. That is because of stochastic fluctuations, associated with turbulence in the underlying accretion
flow in the GRMHD simulations that produce large variations in image structure. On the other hand, data products are calibrated and self-calibrated. There is, therefore, dominance of the stochastic image features over the observational noise. This implies that individual snapshots are highly unlikely to provide an acceptable fit with $\chi_\nu^2 \approx 1$. This problem is solved as follows. An average snapshot image from an ensemble of snapshot images is generated and fitted to data by \texttt{THEMIS} Average Image Scoring (AIS). That is, models are rejected if none of the snapshots are as similar to the average image as the data. In the sixth letter it is noted that “the ‘true’ model is necessarily accepted by the \texttt{THEMIS}-AIS procedure” \cite{EHTCf}, p. 34. That is, the model that would fit the evidence would be the “true” model.

It is shown in the fifth letter that the distribution of $\frac{M_{Bh}}{d}$ from fitting snapshot images to data using \texttt{THEMIS} and \texttt{GENA} gives qualitatively similar results. As better fits are required, the distribution of $\frac{M_{Bh}}{d}$ from fitting snapshot images to 2017 EHT data narrows and peaks close to  $\frac{M_{Bh}}{d} \approx 3.6 \mu as$. So, most models favor $\frac{M_{Bh}}{d} \approx 3.6 \mu as$. The exception is the $a_*=-0.94$ and $R_{high}=1$ SANE model which favors a small $\frac{M_{Bh}}{d} \approx 2 \mu as$ \cite{EHTCe}, p. 10.

The best fit $PA$ were then searched for. Recall that $PA$ approximately corresponds to the orientation of the black hole spin $a_*$. The SANE and MAD snapshot images were divided into two groups: the spin-away models\footnote{In which the black hole’s spin $a_*$ points away from Earth ($i>90^0$, and $a_* >0$, or $i < 90^0$ and $a_* <0$).} and the spin-toward models\footnote{In which the black hole’s spin $a_*$ points toward Earth ($i>90^0$, and $a_* <0$, or $i < 90^0$ and $a_* >0$).}. In each group, the accretion flow either moves with the black hole’s spin or against it (prograde and retrograde, respectively).\footnote{If the accretion flow’s angular momentum and that of the black hole $J$ are aligned, the accretion disk is prograde ($a_* \geq 0$) with respect to the black hole spin axis. But if the black hole’s angular momentum $J$ is opposite that of the accretion flow, the accretion disk is retrograde ($a_*<0$).} 

The values of the large-scale jet $PA$ were found by the fitting procedure (\texttt{THEMIS} and \texttt{GENA}) to be $150^0$-$200^0$ east to north, consistent with the spin-away models and inconsistent with the spin toward models. The two chosen cases – prograde ($i= 163^0$ and $a_*>0$) and retrograde ($i=17^0$ and $a_*<0$) – were therefore the spin-away models, the ones in which the spin of the black hole $a_*$ must always be moving clockwise (as seen from Earth) and the bright section is at the bottom part of the ring. This means there is a persistent asymmetry with the brightest region to the South \cite{EHTCe}, p. 8, p. 10.

It was found that the large-scale jet $PA$ lies on the shoulder of the spin-away models but lies off the shoulder of the spin-toward models. The conclusion was: the alternative GRMHD model images having a bright section of the ring at the top – the spin-toward models, $a_*$ is moving anticlockwise (as seen from Earth) – do not represent observations. The snapshot images with the bright section of the ring at the bottom capture the qualitative features found in the 2017 EHT 2017 April 6 image. 

It is explained in the fifth letter that the ring is brighter at the bottom because the plasma is moving toward us. Due to Doppler beaming, at the top, the ring is less bright because the material is moving away from us. While the approaching side of the plasma of the forward jet is Doppler boosted, the receding side is Doppler dimmed, producing a surface brightness contrast \cite{EHTCe}, p. 3. 

The above finding implies that the sense of rotation of both the jet and the funnel wall (the strongly magnetized region of the accretion flow) are controlled by the black hole spin $a_*$ \cite{EHTCe}, p. 8. This was a hint that the Blandford-Znajek process may be confirmed here \cite{Blandford and Znajek}. More on this below.
\vspace{1mm} %1mm vertical space

After performing the AIS test, only very few models were rejected. In the fifth letter it is said that “the majority of the simulation library models is consistent with the data” and it is then explained that “Given the uncertainties in the model – and our lack of knowledge of the source
prior to EHT2017 – it is remarkable that so many of the models are acceptable” \cite{EHTCe}, pp. 5-15, p. 19; \cite{EHTCf}, p. 10, pp. 14-15, pp. 30-33; \cite{Broderick1}, p. 4.

What that meant was that the majority of the models were “true” models. The ensuing steps therefore required narrowing down the range of the best-fit models by imposing three constraints, the most important of which was the jet power constraint. Based on measurements performed in 2012, 2015 and 2016, the jet power $P_{jet}$ of the core of M87 was assumed to be large: $P_{jet}>10^{42} ergs^{-1}$. The jet power constraint rejected the largest number of models. 

Although in the GRMHD models, “the most likely value from the stellar absorption-line work $M_{BH} \approx 6.2 \times 10^9 M\odot$" was used, for some GRMHD models, images were also generated with $M_{BH} \approx 3.5 \times 10^9 M\odot$ “to check that the analysis results are not predetermined by the input black hole mass” \cite{EHTCe}, p. 5.
\vspace{1mm} %1mm vertical space

For a given magnetic field configuration: $P_{jet} = \frac{\dot{M}c^2}{10}$. 
\vspace{1mm} %1mm vertical space
It was found by \texttt{THEMIS}-AIS that the SANE model with $a_*-0.94$ and $R_{high}=1$ and with $M_{BH} \approx 3.5 \times 10^9 M\odot$ (and $\frac{M_{Bh}}{d} \approx 2 \mu as$) fitted the data. 

But with this value of $M_{BH}$, $\dot{M}$ drops by a factor of two and so $P_{jet}<10^{42} ergs^{-1}$. Consequently, the SANE model with $a_*-0.94$ and $R_{high}=1$ was finally rejected. So, eventually the conclusion was: it is unlikely that the 2017 data capture an $a_*-0.94$ and $R_{high}=1$  SANE model, namely, it is unlikely that the 2017 EHT data capture a $M_{BH} \approx 3.5 \times 10^9 M\odot$ black hole. 

Moreover, for SANE and MAD models that produced sufficiently powerful jets and were consistent with the 2017 EHT data, $P_{jet}$ was found to be driven by the extraction of black hole spin energy through the Blandford-Znajek process.\footnote{In 1969 Roger Penrose suggested that rotational energy can be extracted from the rotating Kerr black hole \cite{Penrose}, p. 1160. In 1977, Roger Blandford and Roman Znajek extended Penrose’s “mechanical extraction of energy” process to electromagnetic extraction of energy from a Kerr supermassive black hole in a MHD environment \cite{Blandford and Znajek}, p. 434, p. 451.} The Kerr black hole is rotating like a conductor in a magnetic field and the total jet power is:

\begin{equation} \label{equation6}
P_{jet} = \frac{k}{4 \pi} \Phi_{BH}^2 \Omega^2,
\end{equation}
  
\vspace{1mm} %1mm vertical space
\noindent where $k \approx 0.045$ is a numerical constant which depends on the magnetic field, $\Phi_{BH}$ is given by equation (\ref{equation2}),   $\Omega^2$ is the angular velocity of the horizon:
\vspace{2mm} %2mm vertical space

\noindent $\Omega^2= \frac{a_*c}{2{r}_+}$ and $r_+$ is given by equation (\ref{equation3}). 
\vspace{2mm} %2mm vertical space

$P_{jet}$ increases quadratically with both the black hole spin $a_*$ and the magnetic flux $\Phi_{BH}^2$, where the spin $-1<a_*<1$. Hence, GRMHD models with $P_{jet}<10^{42} ergs^{-1}$ and zero spin $a_*=0$ are
outright rejected. For $a_*=0$ the bright section of the ring creates a nearly symmetric ring, which is inconsistent with what is seen in the 2017 EHT consensus image \cite{EHTCe}, pp. 9-10, p. 13-15; \cite{Broderick2}, pp. 2-3.

\section{The Kerr black hole hypothesis and alternative hypotheses} \label{Section 5} 

\textbf{The angular gravitational radius and ring diameter.} The next stage of the experiment was obtaining the angular gravitational radius $\theta_g$ by three different methods – GRMHD model fitting, geometric model fitting and image domain feature extraction; and estimating the ring diameter $\theta_d$ by the two latter methods:
\vspace{1mm} %1mm vertical space

1) \emph{Geometric model fitting}: Two kinds of geometric crescent models (called \emph{xs-ring} and \emph{xsringauss}) – collectively named, the \emph{generalized crescent model} (GC) – were developed and compared directly with the 2017 EHT data. One model was compared with data using the \texttt{dynesty} Python Bayesian dynamic NS sampling code while the other was fitted to data by \texttt{THEMIS}. \footnote{NS sampling is an Approximate Bayesian Computation method. In Bayesian statistics, we start with Bayes’ theorem and update the prior probability distribution of the model parameters upon receiving new data to obtain the posterior probability distribution of the parameters. NS is designed to evaluate the Bayesian evidence $\mathcal{Z}$ but as a by-product it can further sample the posterior probability distribution. $\mathcal{Z}$ is calculated by sampling nested points. $N$ live points are sampled from the prior space. At each iteration, the live point with the lowest likelihood $\mathcal{L}_i$ among the live points is found and a new live point is sampled. If $\mathcal{L}_{i+1} \geq \mathcal{L}_i$, the old live point with $\mathcal{L}_i$ is rejected but its values are stored to calculate $\mathcal{Z}$. Dynamic NS (\texttt{dynesty}) has been developed to increase the accuracy of nested sampling, to sort likelihoods more efficiently than NS and speed up the process \cite{EHTCf}, p. 4, p. 6.} Both \texttt{THEMIS} and \texttt{dynesty} produced the following mean value for the ring diameter of the two geometric crescent models: $\theta_d \approx 43 \mu as$.

The GC model does not provide any scientific explanation for the 2017 EHT data because no underlying mechanism based on physics is responsible for the structures in it. The parameters of the GC model are therefore calibrated, i.e. fitted, to the parameters of the GRMHD models. The crescent models are associated with the emission surrounding the shadow of a black hole. If the geometric crescent is formed by gravitational lensing, the angular diameter of the photon ring obeys the equation:

\begin{equation} \label{equation7}
\theta_d= \alpha \theta_g.
\end{equation}

\noindent $\alpha$ represents the gravitational lensing factor and $\theta_g$ is defined by equation (\ref{equation5}). \texttt{dynesty} produced a ring diameter of $\theta_d \approx 43.2 \mu as$ for the \emph{xs-ring} model and \texttt{THEMIS} produced $\theta_d \approx 43.4 \mu as$ for the \emph{xs-ringauss} model. Those values were then compared with the known value of $\theta_g$ that went into the GRMHD simulations.\footnote{This led to mean values of $\alpha$: $\alpha=11.55$ for the \emph{xs-ring} model and $\alpha=11.50$ for the \emph{xs-ringauss} model. The two $\theta_d$ measurements were combined with the two $\alpha$ values to arrive at values of $\theta_g$ using equation (\ref{equation7}): $\frac{43.2 \mu as}{11.55}$ and $\frac{43.4 \mu as}{11.50}$.} The final inferred value is $\theta_g=3.77_{-0.40}^{+045}\mu as$ \cite{EHTCf}, pp. 10-13.
\vspace{2mm} %2mm vertical space
  
2) \emph{GRMHD model fitting}: GRMHD snapshot images were fit directly to data by \texttt{THEMIS} and \texttt{GENA} (see Section \ref{Section 3}). The combined value for the analysis performed by both \texttt{THEMIS} and \texttt{GENA} is: $\theta_g=3.80_{-0.31}^{+039}\mu as$.
\vspace{1mm} %1mm vertical space

3) \emph{Image domain feature extraction}: The values of $\theta_d$ obtained by image domain feature extraction (see end of Section \ref{Section 2}) were converted to the $\theta_g$ using the scaling factor $\alpha$ [Equation (\ref{equation7})], following a similar procedure to the one used in the geometric model fitting. The following value of $\theta_g$ was obtained: $\theta_g=3.83_{-0.36}^{+042}\mu as$. 
\vspace{3mm} %3mm vertical space

Fitting geometric models to data and extracting feature parameters in the image domain both allowed quantifying the following properties of the object at the center of M87: an angular ring diameter of $\theta_d \approx 42 \pm 3 \mu as$, an angular gravitational radius of $\theta_g \approx 3.8 \pm 0.4 \mu as$, a deep central brightness depression (the fractional central brightness: the ratio of the mean brightness interior to the ring to the mean brightness around the ring), and $PA$. In the sixth letter it is concluded: “All of these features support the interpretation that we are seeing emission from near the event horizon
that is gravitationally lensed into a crescent shape near the photon ring” \cite{EHTCf}, pp. 20-21.

Prior measurements based on stellar dynamics had produced the value $\theta_g=3.62_{-0.34}^{+060}\mu as$. This value is consistent with the values calculated by the EHT project. On the other hand, gas dynamics measurements led to a lower value: $\theta_g=2.05_{-0.16}^{+048}\mu as$.
\vspace{1mm} %1mm vertical space

It is remarked in the sixth letter that “All of the individual $\theta_g$ estimates use the GRMHD simulation library, either through directly fitting GRMHD snapshots to the data […GRMHD model fitting] or through calibration of diameters [$\theta_g$] resulting from geometric models or reconstructed images”, i.e. from geometric crescent models or image domain feature extraction. Hence “A degree of caution is therefore warranted. The measurements rely on images generated from GRMHD simulations and should be understood within that context” \cite{EHTCd}, pp. 27-30; \cite{EHTCe}, p. 2; \cite{EHTCf}, p. 1, pp. 4-21, p. 31, p. 39.
\vspace{1mm} %1mm vertical space

\textbf{The mass of the black hole.} In the first letter it is noted that “A basic feature of black holes in GR is that their size scales linearly with mass” \cite{EHTCa}, p. 9. The estimation of the angular ring diameter of $\theta_d \approx 42 \pm 3 \mu as$ should therefore allow for the determination of the mass $M_{BH}$ of the core of M87. 

In 1973, James Maxwell Bardeen found that the black hole casts a shadow on the hot gas that surrounds it. The diameter of the apparent shadow $D_{sh}$ seen by a distant observer depends on the gravitational lensing $\alpha$ around the black hole. For a Schwarzschild black hole: $D_{sh}=2 \sqrt{27} R_g \approx 10.39 R_g$ where $\alpha=2 \sqrt{27}$. However, in the vicinity of a rotating Kerr black hole, “the effect of the frame dragging induced by the angular momentum of the Kerr black hole is quite apparent” \cite{Bardeen}, pp. 230-233.\footnote{The Lense-Thirring effect due to the black hole rotation acts to compress the shadow with respect to the rotation axis while the quadruple moment of the rotating black hole causes an oblate shape of the shadow. The two effects approximately cancel each other out and we are left with a nearly circular shadow.} Thus, for a Kerr black hole:

\begin{equation} \label{equation8}
D_{sh}<2 \sqrt{27} R_g \approx 9.8 R_g, R_{sh}<\sqrt{27} R_g \approx 4.9 R_g.
\end{equation}
  
\vspace{1mm} %1mm vertical space
\noindent In the GRMHD images: $D_{sh}=9.79 M_{BH}$.

The radius $R_g$ is given by equation (\ref{equation4}). Combining equation (\ref{equation8}) with equation (\ref{equation5}) gives:

\begin{equation} \label{equation9}
\theta_d \approx \frac{2\sqrt{27} R_g}{d} \approx \frac{9.8 R_g}{d} \equiv \frac{9.8 G M_{BH}}{d c^2} = \alpha \theta_g.
\end{equation}

\vspace{1mm} %1mm vertical space
Inserting $\theta_d \approx 42 \pm 3 \mu as$ and $ d = 16.8 \pm 0.8$ Mps into equation (\ref{equation9}) gives $M_{BH}$. It was shown that $\theta_d$ is consistent with a mass of $M_{Bh} \approx 6.5 \pm 0.7 \times 10^9 M\odot$. 

Recall that measurements based on stellar dynamics had produced the value: $M_{Bh} \approx 6.2 \times 10^9 M\odot$ \cite{Gebhardt}. This estimation is consistent with the one calculated by the EHT project. On the other hand, gas dynamics measurements had led to a lower value of $M_{Bh} \approx 3.5 \times 10^9 M\odot$ \cite{Walsh}, see Section \ref{Section 4}. 

\vspace{1mm} %1mm vertical space
In the fifth letter it is stressed: “Although our working hypothesis has been that M87 contains a Kerr black hole, it is interesting to consider whether or not the data is also consistent with alternative models for the central object” \cite{EHTCe}, p. 17.

One such alternative model is a black hole that is spinning more rapidly than the Kerr bound, defined as follows: $a_* \leq M_{BH}$ or $J\leq M_{BH}^2$ ($J\leq \frac{GM_{BH}^2}{c}$) see equations (\ref{equation1}) and (\ref{equation3}). Those black holes are super-spinning black holes and are called superspinars. Their horizon would disappear if $J=a_*$ and this would imply the existence of a naked singularity and the violation of causality (spinning faster than the speed of light). Computer simulations show that the shadows of superspinars are significantly smaller compared with those of Kerr black holes \cite{EHTCe}, p. 18.

Hence, there exist solutions of the field equations of general relativity which describe naked singularities not hidden by an event horizon. But the possible existence of these exotic solutions causes severe problems to the uniqueness theorems. The reason is that only if all singularities are surrounded by event horizons, the Kerr black hole is a unique solution. So, in 1969 and 1976 Roger Penrose and Stephen Hawking phrased the censorship hypothesis, which says that physics censors naked singularities by always enshrouding them with a horizon \cite{Penrose},  p. 1160, p. 1162; \cite{Hawking}, p. 2461.

To verify the censorship hypothesis, we should demonstrate that the core of M87 is not a naked singularity/superspinar. There also exist other horizonless solutions of the field equations such as rotating wormholes. According to equation (\ref{equation8}), for a Kerr black hole: $R_{sh} \approx 4.9 R_g$. A $6.5 \times 10^9 \odot$ rotating wormhole has half as big a shadow radius as a $6.5 \times 10^9 \odot$ black hole: $R_{sh} \approx 2.7 R_g$. This is about the same size as the shadow of the $3.5 \times 10^9 \odot$ black hole. Hence, for a $3.5 \times 10^9 \odot$ black hole we would get a smaller shadow than for a $6.5 \times 10^9 \odot$ black hole. A $6.5 \times 10^9 \odot$ naked singularity (superspinar) has an even smaller shadow radius of $R_{sh} = R_g$. 

The different shadow sizes ($R_{sh}$) were then overlaid on top of the April 11 2017 EHT fiducial images from each imaging pipeline (the consensus image). The exotic possibilities were ruled out and one was left with a ring that nearly perfectly matches that of a $6.5 \times 10^9 \odot$ black hole \cite{EHTCa}, pp. 8-9; \cite{EHTCe}, p. 18; \cite{Johnson}. 

But there is a fly in the ointment because demonstrating that the core of M87 has an event horizon does not conclusively disprove the existence of naked singularities elsewhere \cite{Psaltis and Doeleman}, p. 77.   
 
In the first letter it is remarked that “However, other compact-object candidates need to be analyzed with more care”. And in the fifth letter it is again stressed: “Future observations and more detailed theoretical modeling, combined with multiwavelength campaigns and polarimetric measurements, will further constrain alternatives to Kerr black holes”. It is further noted that “the comparisons carried out here must be considered preliminary. Nevertheless, they show that the EHT2017 observations are not consistent with several of the alternatives to Kerr black holes” \cite{EHTCa}, pp. 8-9; \cite{EHTCe}, p. 18.

\section{The Kerr black hole is the most likely explanation of the evidence} \label{Section 6} 

A null hypothesis test, which consists of three ingredients, was subsequently performed:
\vspace{1mm} %1mm vertical space

1) The dynamical measurements provide an accurate determination of $M_{BH}$. 

2) The shadow in the 2017 EHT image is a black hole shadow. 

3) The space-time of the black hole is described by the Kerr solution.

\vspace{2mm} %2mm vertical space

\emph{Ingredient 1}: An estimation was performed of the difference between the measurement of $\theta_g$ performed by the EHT Collaboration and prior measurements of $\theta_{dyn}$:

\begin{equation} \label{equation10}
\delta \equiv \frac{\theta_g}{\theta_{dyn}} - \frac{\theta_g}{\theta_g}= \frac{\theta_g}{\theta_{dyn}} - 1. 
\end{equation}

\vspace{1mm} %1mm vertical space
For gas dynamics: $\delta \approx 0.78$ and for stellar dynamics: $\delta \approx -0.01$. In the sixth letter it is written that the fact that the EHT measurement of $\theta_g$ is consistent with one of the measurements “allows us to conclude that our null hypothesis has not been violated” \cite{EHTCf}, pp. 22-23.
\vspace{1mm} %1mm vertical space

\emph{Ingredient 2}: Are black holes described entirely by their mass $M_{BH}$ and spin $a_*$, or do they have hair (are they described by other parameters)? To answer this question, we have to know $M_{BH}$ because the size of the shadow is proportional to $M_{BH}$. Violations of the above no-hair theorem generically change the shape and size of the shadow of the black hole. In other words, for Kerr black holes of known $M_{BH}$, the size and shape of the shadow remain nearly unchanged and shadows always appear nearly circular. Detecting a shadow and extracting its characteristic properties offers a chance to constrain the space-time metric. For a black hole of known mass $M_{BH}$ and distance $d$ from Earth, identifying the presence of the shadow and confirming that its size is in the narrow range $\left(4.9-5.2 \right)M_{BH}$ constitutes a statistical null hypothesis test of the Kerr black hole \cite{Psaltis and Doeleman}, pp. 77-79; \cite{Psaltis}, p. 1; \cite{EHTCf}, p. 2; \cite{Johnson}. As said at the end of section \ref{Section 5}, different $R_{sh}$ were overlaid on top of the consensus image. The ring of the 2017 EHT image nearly perfectly matches that of an $R_{sh} \approx 4.9 M_{BH}$ black hole whose mass is $M_{BH} \approx 6.5 \pm 0.7 \times 10^9 M\odot$. 
\vspace{1mm} %1mm vertical space

\emph{Ingredient 3}: In the fifth letter, it is explained: “we now adopt the working hypothesis that M87 contains a turbulent, magnetized accretion flow surrounding a Kerr black hole. To test this hypothesis quantitatively against the EHT2017 data we have generated a Simulation Library of 3D
time-dependent ideal GRMHD models” \cite{EHTCe}, p. 3. And: “all [GRMHD] models assume a Kerr black hole space-time, but there are alternatives” \cite{EHTCe}, p. 15. It is shown that GRMHD models that assume a Kerr metric fit the 2017 EHT observations (see Section \ref{Section 3}) and it is concluded that the space-time of the black hole is described by the Kerr metric.
\vspace{2mm} %2mm vertical space

Finally, in the fifth letter, three images are placed one next to the other: a snapshot image based on a (spin-away) GRMHD model of a Kerr black hole, a snapshot image based on a (spin-away) GRMHD model of a Kerr black hole (restored with a FWHM Gaussian $20 \mu as$ beam), and the 2017 EHT April 6 image. One can see that the two blurred images are almost the same \cite{EHTCa}, p. 6; \cite{EHTCe}, p. 2, pp. 5-6, p. 8, p. 10, p. 15. 

\section{Testing the Kerr black hole hypothesis with gravitational waves} \label{Section 8}

It is written in the first letter of the EHT Collaboration: "Altogether, the results derived here provide a new way to study compact-object spacetimes and are complementary to the detection of gravitational waves from coalescing stellar-mass black holes with LIGO/Virgo \cite{EHTCa}, p. 9. I shall now discuss the detection of gravitational waves with Laser Interferometer Gravitational Wave Observatory (hereafter LIGO)/Virgo interferometer.

A paper, “Testing the No-Hair Theorem with GW150914” \cite{Isi}, was published at about the same time as the six letters  published in \emph{The Astrophysical Journal Letters} \cite{EHTCa}-\cite{EHTCf}. The paper presents an analysis of gravitational-wave data from the first LIGO detection of the binary black-hole merger GW150914.\footnote{The gravitational-waves signals are surrounded by noise. Gravitational-wave signals are extracted from the background noise using statistical significance tests and "templete waveforms" (model) fitting to gravitational-wave data. The gravitational waves from the merger GW150914 were so loud that they were visible in the data even with minimal data processing.} 

Two stellar-mass black holes that merge, form a single distorted compact object that gradually settles to a final stationary form. Gravitational waves are emitted throughout the entire process, at each moment carrying information about the evolving compact object. There are three stages in the coalescence of the two black holes:
\vspace{1mm} %1mm vertical space

1) \emph{Inspiral}: a long phase in which the two black holes are still quite far one from each other but are slowly orbiting one another in a quasi-circular orbit. Since we are dealing with weak gravitational fields, this phase allows an analytic description using the post-Newtonian expansion (approximation). During this phase, the amplitude and frequency of the gravitational waves are increasing.   

2) \emph{Merger}: the distance between the two black holes gets smaller, the orbit gradually shrinks and the two bodies finally merge. Numerical relativity (numerical simulations) is required. That is because the gravitational field is so strong and also time-dependent. During this phase, the amplitude of the gravitational waves gradually increases until it reaches a maximum.   

3) \emph{Ringdown}: the newly created object releases its final gravitational wave signals away into endless space. It wobbles and oscillates during which it rings like a bell with characteristic frequencies and damping times determined entirely by the mass and spin of the black hole. This stage is treated analytically using perturbation theory. 
A linearly perturbed Kerr black hole emits gravitational waves in the form of exponential damped sine waves, with specific frequencies and decay rates determined exclusively by the black hole’s mass and spin. 

The ringdown phase consists of a superposition of quasi-normal modes. Each quasi-normal mode has a characteristic complex angular frequency: the real part is the angular frequency and the imaginary part is the inverse of the damping time. These modes are distinguished by their longitudinal and azimuthal indices, $l$ and $m$ respectively, as well as by their overtone number $n$. Ringdown overtones are the quasi-normal modes with the fastest decay rates. Each mode has a particular frequency and decay rate which are functions of the Kerr black hole parameter spin and total mass of the black hole that is being perturbed. Numerical simulations have demonstrated that the fundamental mode $l = m = 2$ seems to dominate the ringdown signal \cite{Buonanno}, p. 124018-18, p. 124018-21.
\vspace{1mm} %1mm vertical space

In 2016, the LIGO Scientific and Virgo Collaborations announced the first joint detection of gravitational waves (GW150914) with both the LIGO and Virgo detectors. In 2018, they reported of several quantitative tests made on the gravitational-wave data from the detection of GW150914. The purpose of the tests was to validate GR. The result of the tests indicates that the entire GW150914 inspiral-merger-ringdown (IMR) waveform does not deviate from the predictions of a binary black-hole merger in classical general relativity \cite{LIGO}, pp.221101-3-221101-5.  

The detected gravitational-wave signal increases in frequency and amplitude in about eight cycles from $35$ to $150$ Hz, where it reaches a peak amplitude. After a time around $0.42 s$, the amplitude drops rapidly, and the frequency appears to stabilize. After the peak gravitational wave amplitude is reached, the signal makes one to two additional cycles, continuing to rise in
frequency until reaching about $250$ Hz, while dropping
sharply in amplitude. 

The most plausible explanation for this empirical evolution is gravitational-waves emission from two orbiting masses in the inspiral and merger phases. The drop in amplitude is consistent with a Kerr black hole. Recall that for a Kerr black hole, the ringdown is expected to have a damping time roughly equal to the period of oscillation (inversely equal to the frequency).
\vspace{1mm} %1mm vertical space
For a black hole with spin $\chi = 0.7$, a ringdown frequency of $\approx 260$ Hz$\left(\frac{65M\odot}{M}\right)$ and a damping time: $4ms\left( \frac{M}{65M\odot}\right)$ are calculated. 
\vspace{1mm} %1mm vertical space
The LIGO Scientific and Virgo Collaborations conclude: "the signal in the data is fully consistent with the final object being a Kerr black hole with a dimensionless spin parameter $\chi = 0.7$ and a mass $65M\odot$" \cite{LIGO2}, pp. 2-3, pp. 12-13.  

The LIGO Scientific and VIRGO Collaborations explain that the initial behavior of the gravitational-wave signal "cannot be due to a perturbed system returning back to stable equilibrium, since oscillations around equilibrium are generically characterized by roughly constant frequencies and decaying amplitudes". On the other hand, the GW150914 data "demonstrate very different behavior. During the period when the gravitational wave frequency and amplitude are increasing, orbital motion of two bodies is the only plausible explanation: there, the only 'damping forces' are provided by gravitational wave emission, which brings the orbiting bodies closer (an 'inspiral'), increasing the orbital frequency and amplifying
the gravitational wave energy output from the system" \cite{LIGO2}, p. 3.

It is concluded: "our inspiral–merger–ringdown test shows no evidence of discrepancies with the predictions of GR" \cite{LIGO}, p.221101-5. 

The LIGO Scientific and Virgo Collaborations performed a test that allowed "for possible violations of GR" and concluded that the value of parameters was usually found to represent GR. An additional test was performed. They asked: Does GR fit the data better than alternative competing models? The following model was compared with GR: a model based on theories of gravity mediated by a graviton with a nonzero mass in which the gravitational waves travel at a speed different than the speed of light. In GR, gravitons are massless and travel at the speed of light. The presence of non-GR polarization states were searched for. But it was difficult to distinguish between GR and non-GR models on the basis of GW150914 data alone. Since the Hanford and Livingston LIGO instruments have similar orientations, they are sensitive to a very similar linear combination of the gravitational-wave polarizations, so it is difficult to distinguish between the GR and non-GR states \cite{LIGO}, p.221101-10. 

It is concluded: "With the exception of the graviton Compton wavelength and the test for the presence of a non-GR polarization, we did not perform any studies aimed at constraining parameters that might arise from specific alternative theories" \cite{LIGO}, p. 221101-8, pp. 221101-10-11. 
\vspace{1mm} %1mm vertical space

Testing the black hole ring at the correct frequencies and damping times, it ought to be possible to test the validity of the no-hair theorem, which states that the remnant must be a Kerr black hole.

Already in 2004, a team of astrophysicists had asked: “can gravitational wave observations provide a test of one of the fundamental predictions of general relativity: the no-hair theorem?” They suggested “a definitive test of the hypothesis that observations of damped, sinusoidal gravitational waves originate from a black hole or, alternatively, that nature respects the general relativistic no-hair theorem” \cite{Dreyer}, p. 787. 

In 2017, three astrophysicists belonging to the LIGO Scientific and Virgo Collaborations, Eric Thrane, Paul Lasky and Yuri Levin, wrote that “The recent detections of gravitational waves from stellar-mass black hole mergers would seem to suggest that a test of the no-hair theorem might be around the corner”. A method was provided "for testing the no-hair theorem using only data from after the remnant black hole has settled into a perturbative state”. 

The post-merger remnant must be allowed to settle into a perturbative, Kerr-like state and this means that the ringdown frequencies and damping times depend only on the mass and spin of the newly created black hole. In this way, the no-hair theorem places stringent requirements on the asymptotic behavior of perturbed black holes. The no-hair theorem concerns itself with linear perturbations. But it was found that at no point in time is the post-merger waveform precisely described by black hole perturbation theory. That is because, there is always a contribution, however small, left over from the merger. Furthermore, the ringdown signal becomes weaker as it settles into a Kerr black hole. 

The authors were confronted with the following problem: they could either obtain higher signal-to-noise (SNR) ratio by adding louder signals from the merger phase (the main gravitational wave signal) or, wait for the remnant black hole to settle to the perturbative state where the no-hair theorem applies. They chose the latter option because as louder signals were added, it was no longer clear whether this improved the knowledge of the ringdown frequency and damping times. So, they arrived at the conclusion that it may be possible to test the no-hair theorem but the observed behavior could either be attributed to a numerical relativity artifact, or to a residual non-linearity of the signal from the merger decaying on a timescale comparable to the linear ringdown signal \cite{Thrane}, pp. 102004-1-2, p. 102004-5.

Matthew Giesler, Mark Scheel and Saul Teukolky have been working on the LIGO data and gravitational waves. In May 2019, they submitted a paper with Maximiliano Isi and Will Farr from the LIGO Scientific and Virgo Collaborations in which they analyzed the ringdown gravitational-wave data. They thought that a gestalt shift was needed. It does not mean that Thrane, Lasky and Levin's reservations are not sound. It only means that one team thinks they have solved the problem whilst the other probably holds an opposite view.  

So, Geisler and his team have shown that including enough overtones of the mode $l = 2$ allows obtaining higher SNR. 
A GW150914-like signal was first studied.\footnote{A simulated output of a LIGO-like detector was studied in response to the same gravitational waves as the ones from the asymptotic remnant. That is, the $l = m = 2$ mode of the signal was injected into simulated Gaussian noise corresponding to the sensitivity of Advanced LIGO in its design configuration.} It was found that the inclusion of enough overtones associated with the $l = m = 2$ mode “provides a high-accuracy description of the ringdown”, where the high SNR “can be exploited to significantly reduce the uncertainty in the extracted remnant properties” \cite{Giesler}, p. 041060-2, p. 041060-8, p. 041060-10. 

Previously, overtones had been believed to be too faint to be detected and it was believed that one had to wait at least ten years to test the no-hair theorem. But as shown above, studies done by the team have shown that although the overtones decay very quickly, the first overtones of GW150914 should be loud enough to be detected. 

Measuring the quasi-normal modes from gravitational-wave observations is tantamount to measuring the ringdown spectrum. The ringdown spectrum is "a fingerprint that identifies a Kerr black hole" and the measurements of this spectrum "has been
called black-hole spectroscopy" \cite{Isi}, p. 111102-1. 

Isi et al. were “Assuming the remnant is a Kerr black hole”. That is, “assuming first that quasinormal modes are as predicted for a Kerr black hole within general relativity”, they checked this assumption \cite{Isi}, p. 111102-2. 

Making use of overtones, Isi et al. extracted information about
the GW150914 remnant using only post-inspiral data, starting at the peak of the signal. The mass and spin of the remnant were measured using the overtones of the mode $l = m = 2$. The GW150914 ringdown using this mode and its first overtone were analyzed. These are ringdown-only measurements. The ringdown-only measurements of the remnant mass and spin magnitude were then compared to those obtained from the analysis of the entire GW150914 inspiral-merger-ringdown (IMR) waveform. It was concluded that evidence of the mode $l = m = 2$ and at least one overtone was found and, a $90 \%$-credible measurement of the remnant mass and spin was obtained in agreement with that inferred from the full waveform. 

Measuring the frequencies of the fundamental and first overtone, information was extracted about the GW150914 remnant and "their consistency with the Kerr hypothesis" was established. It was concluded that the GW150914 merger "produced a Kerr black hole as described by general relativity" \cite{Isi}, p. 111102-5. 

It is stressed at the end of Isi et al.'s paper that, as the precision of the detectors improves, it can be expected that using overtones will provide better tests of the no-hair theorem \cite{Isi}, p. 111102-1, p. 111102-3, p. 111102-5. 

\section{Inference to the best explanation} \label{Section 7}

The first question asked by the EHT Collaboration is: Does M87’s core have a horizon? Is the object at the center of the galaxy M87 a spinning Kerr black hole as predicted by general relativity? 

In order to answer this question, the EHT Collaboration first endorsed the working hypothesis that the central object is indeed a black hole described by the Kerr metric. They chose this hypothesis based on previous research and observations of the galaxy M87. In 2015 a team of scientists belonging to the EHT Collaboration made an experiment the goal of which was to compare the merits of two rival hypotheses: a compact object with a surface and a black hole with a horizon. The black hole hypothesis was selected as the best explanation of the evidence. 

The EHT Collaboration tested the Kerr black hole hypothesis and found that it was consistent with the evidence. How did they test the Kerr black hole hypothesis? GRMHD models based on the Kerr metric were used to extract parameters from the 2017 EHT data (see Sections \ref{Section 3}, \ref{Section 4} and \ref{Section 5}) and then it was shown that the data confirm the hypothesis that the compact object at the center of M87 is a supermassive Kerr rotating black hole (see Section \ref{Section 6}). The sixth letter ends with the following statement: “Together, our results strongly support the hypothesis that the central object in M87 is indeed a Kerr black hole” \cite{EHTCf}, p. 23.

Ian Hacking once said: “This looks like a classic example of what philosophers, following Gilbert Harman (\cite{Harman}) have come to call inference to the best explanation (IBE)”. He added: “there is an overall conjecture of the sort that Popper would have called metaphysical” (\cite{Hacking 1989}, pp. 568-569). In our case study, in the fifth letter it is written: “the results provided here are consistent with the existence of astrophysical black holes” (\cite{EHTCe}, p. 19). Hacking and Wesley Salmon are on the same page when it comes to supermassive black holes.\footnote{Salmon wrote: astrophysicists believe that there is a black hole at the center of our galaxy. But he was not convinced of this particular existence claim and he added that Hacking was “skeptical far more generally about black holes”. From what emerges from Hacking’s papers, he is indeed skeptical about black holes.} They have argued that scientific existence claims are always open to question. When one says that astrophysical black holes exist, it adds nothing to our understanding and to the explanation. The fact that the black hole hypothesis explains the observations does not provide evidence that black holes exist. Existence claims with respect to black holes are not part of the explanation \cite{Hacking 1983}, pp. 52-54; \cite{Salmon}, p. 56, p. 58.

The Kerr black hole hypothesis has explanatory power in accounting for: \emph{ingredient 2}, \emph{ingredient 3} (see Sections \ref{Section 6} and \ref{equation3}) and the GW150914 Ringdown. It is also the best explanation of the evidence for M87's core and for the GW150914 gravitational wave event. Does it mean that it is also true? If truth means that no other hypothesis will explain the evidence as well as the Kerr black hole hypothesis, then the answer is: we cannot say that it is true. The reason is that the EHT Collaboration has written: “Future observations and more detailed theoretical modeling, combined with multiwavelength campaigns and polarimetric measurements, will further constrain alternatives to Kerr black holes” \cite{EHTCe}, p. 18; see the end of Section \ref{Section 5}. And Isi et al. write: "Future studies of black-hole ringdowns relying on overtones could potentially allow us to identify black-hole mimickers", that is, exotic alternatives to the Kerr black hole hypothesis, "and probe the applicability of the no-hair theorem with high precision, even with existing detectors". This would "lead to more specific predictions from general relativity" \cite{Isi}, p. 111102-5.

\section{Underdetermination of theory by evidence} \label{Section 9}

Doeleman has said: "Now if we could make an image of this object, then we could test Einstein’s theories of gravity in the one place where they might break down in the universe” \cite{Doeleman}. Can we also test Hawking radiation in the one place where Einstein's theory of gravity breaks down? 

Hawking pointed out that evaporation violates the classical censorship hypothesis. Hawking explained that if one tries to describe the process of a black hole losing mass and eventually disappearing and evaporating by a classical space-time metric, then “there is inevitably a naked singularity when the black hole disappears. Even if the black hole does not evaporate completely one can regard the emitted particles as having come from the singularity inside the black hole and having tunneled out through the event horizon on spacelike trajectories” \cite{Hawking}, p. 2461; see \cite{Weinstein2}, p. 23.

Since the rate of evaporation of a supermassive black hole is so slow and the Hawking radiation is too faint to be observable, the 2017 EHT observations can neither confirm nor refute Hawking’s evaporation. And since the evaporation time of a supermassive black hole is no less than $10^{94}$ years, then according to Hawking’s hypothesis at the time at which the EHT experiment was performed in 2017, the compact object at the center of M87 could not be a naked singularity.\footnote{The evaporation time $t$ for the black hole, from the point of view of an outside observer, is proportional to the cube of the mass of the black hole: $t=\frac{5120 \pi G^2 M_{BH}^3}{\hbar c^4}$ \cite{Weinstein2}, p. 18.} And according to Hawking’s theory, it will have a horizon for many years to come. Paradoxically, Hawking's above hypothesis can explain at least as well the evidence as the no-hair theorem. Although according to Hawking, a supermassive black hole will gradually evaporate through Hawking radiation, the rate of this evaporation is so slow that we cannot detect it. Collecting additional and more accurate data from the object at the center of M87 will not be enough to detect Hawking radiation. 

That is exactly the point raised by Bas van Fraassen: “Is the best explanation we have, likely to be true?” He answers this question in the negative saying that there are many possible theories, perhaps never yet formulated, but that can fit all the evidence so far. These theories can explain at least as well the evidence as the best theory we have now. But with respect to statements that go beyond the evidence we possess today these theories disagree with one another. We consider them to be false because of underdetermination of theory by evidence. Classical general relativity and the no-hair theorem are our best explanation and they belong to the class of explanations that can fit all the evidence to date, but they also disagree with certain quantum gravity theories, theories with quantum corrections to classical general relativity \cite{Fraassen}, pp. 160-161.


\begin{thebibliography}{34}

\bibitem[1]{Bardeen} Bardeen, J. M. (1973). "Timelike and Null Geodesics in the Kerr Metric." In DeWitt C. and DeWitt
B. S. (eds.). \emph{Black Holes Les Astres Occlus.} New York: Gordon and Breach Science Publishers, pp. 216-239.

\bibitem[2]{Blandford and Znajek} Blandford, R. D. and Znajek, R. L. (1977). “Electromagnetic extraction of energy from Kerr black holes.” \emph{Monthly Notices of the Royal Astronomical Society} 179, pp. 433-456.

\bibitem[3]{Buonanno} Buonanno, A., Cook, G. B., and Pretorius, F. (2007). "Inspiral, merger, and ring-down of equal-mass black-hole binaries." \emph{Physical Review D} 75, pp. 124018-1-124018-42.

\bibitem[4]{Broderick1} Broderick, A. E., Gold, R. G., Karami, M., et al. (Event Horizon Telescope Collaboration) (2020). “Themis: A parameter estimation framework for the event horizon telescope.” \emph{The Astrophysical Journal} 897, pp. 1–38.

\bibitem[5]{Broderick2} Broderick, A. E., Narayan, R., Kormendy, J., Perlman, E. S., Rieke, M. J., and Doeleman, S. S. (2015). “The event horizon of M87.” \emph{The Astrophysical Journal} 805, pp. 1-9.

\bibitem[6]{Doeleman} Doeleman, S. S. (2020). “PSW 2421. The Event Horizon Telescope. First Ever Images of a Black
Hole.” \emph{The 89th Joseph Henry Lecture at the Philosophical Society of Washington Science}.

\bibitem[7]{Dreyer} Dreyer, O., Kelly, B., Krishnan, B. Finn, L. S., Garrison, D., and Lopez-Aleman, R. (2014). "Black-hole spectroscopy: testing general through gravitational-wave observations." \emph{Classical and Quantum Gravity} 21, pp. 787–803.

\bibitem[8]{Falcke} Falcke, H., Melia, F., and Agol, E. (2000). “Viewing the Shadow of the Black Hole at the Galactic
Center.” \emph{The Astrophysical Journal} 528, pp. L13–L16.

\bibitem[9]{Fraassen} van Fraassen, B. C. (1989). \emph{Laws and Symmetry}. Oxford: Oxford University Press.   

\bibitem[10]{EHTCa} Event Horizon Telescope Collaboration (2019). “First M87 event horizon telescope results. I. The
shadow of the supermassive black hole.” \emph{Astrophysical Journal Letters} 875: L1, pp. 1-17.

\bibitem[11]{EHTCb} Event Horizon Telescope Collaboration (2019). “First M87 event horizon telescope results. II. Array and Instrumentation.” \emph{Astrophysical Journal Letters} 875: L2, pp. 1-28.

\bibitem[12]{EHTCc} Event Horizon Telescope Collaboration (2019). “First M87 event horizon telescope results. III. Data processing and calibration.” \emph{Astrophysical Journal Letters} 875: L3, pp. 1-32.

\bibitem[13]{EHTCd} Event Horizon Telescope Collaboration (2019). “First M87 event horizon telescope results. IV. Imaging the central supermassive black hole.” \emph{Astrophysical Journal Letters} 875: L4, pp. 1-52.

\bibitem[14]{EHTCe} Event Horizon Telescope Collaboration (2019). “First M87 event horizon telescope results. V. Physical origin of the asymmetric ring.” \emph{Astrophysical Journal Letters} 875: L5, pp. 1-31.

\bibitem[15]{EHTCf} Event Horizon Telescope Collaboration (2019). “First M87 event horizon telescope results. VI. The shadow and mass of the central black hole.” \emph{Astrophysical Journal Letters} 875: L6, pp. 1-44.

\bibitem[16]{Gebhardt} Gebhardt, K., Adams, J., Richstone, D., Lauer, T. R., Faber, S. M., Gültekin, K., Murphy, J. and Tremaine, S. (2011). “The Black-Hole Mass in M87 from Gemini/NIFS Adaptive Optics Observations.” \emph{The Astrophysical Journal} 729, pp. 1-13.

\bibitem[17]{Giesler} Giesler, M., Isi, M., Scheel, M. A., and Teukolsky, S. A. (2019). "Black Hole Ringdown: The Importance of Overtones." \emph{Physical Review X} 9, pp.  041060-1-041060-13.

\bibitem[18]{Hacking 1983} Hacking, I. (1983). \emph{Representing and Intervening: Introductory. Topics in the Philosophy of Natural Science}. Cambridge: Cambridge University Press.

\bibitem[19]{Hacking 1989} Hacking, I. (1989). “Extragalactic Reality: The Case of Gravitational Lensing.” \emph{Philosophy of Science} 56, pp. 555-581.

\bibitem[20]{Harman} Harman, G. (1965). “The Inference to the best explanation.” \emph {The Philosophical Review} 74, pp. 88-95.

\bibitem[21]{Hawking} Hawking, S. W. (1976). “Breakdown of predictability in gravitational collapse.” \emph{Physical Review} D 14, pp. 2460-2473.

\bibitem[22]{Isi} Isi, M., Giesler, M., Farr, W. M., Scheel, M. A., and Teukolsky, S. A. (2019). "Testing the No-Hair Theorem with GW150914." \emph{Physical Review Letters} 123, pp. 111102-1-111102-6.

\bibitem[23]{Johnson} Johnson, M. D. (2020). “Photographing a Supermassive Black Hole with the Event Horizon Telescope.” \emph {Lecture at Black hole Initiative seminar}, Harvard university, April 12.

\bibitem[24]{LIGO2} LIGO Scientific and Virgo Collaborations (2017). "The basic physics of the binary black hole merger GW150914." \emph{Annalen der Physik} 529, pp. 1-17.

\bibitem[25]{LIGO} LIGO Scientific and Virgo Collaborations (2018). "Tests of General Relativity with GW150914." \emph{Physical Review Letters} 116, pp. 221101-1-221101-19.

\bibitem[26]{Penrose} Penrose, R. (1969). “Gravitational Collapse: The Role of General Relativity.” \emph{Rivista del Nuovo Cimento, Numero Speziale} 1, pp. 252-275; \emph{General Relativity and Gravitation} 34, 2002, pp. 1141-1165.

\bibitem[27]{Porth} Porth, O., Chatterjee, K., Narayan, R., et al. (Event Horizon Telescope Collaboration) (2019). “The Event Horizon General Relativistic Magnetohydrodynamic Code Comparison Project.” \emph{Astrophysical Journal, Supplement Series} 243, pp. 1-40. 

\bibitem[28]{Psaltis and Doeleman} Psaltis, D. and Doeleman, S. (2015). “The Black Hole Test.” \emph {Scientific American} 313, pp. 74-79.

\bibitem[29]{Psaltis} Psaltis, D., Özel, F., Chan, C-K and Marrone, D. P. (2015). “A General Relativistic Null Hypothesis Test with Event Horizon Telescope Observations of the Black Hole Shadow in Sgr A*.” \emph{The Astrophysical Journal} 814, pp. 1-14.

\bibitem[30]{Salmon} Salmon, W. C. (2005). \emph{Reality and Rationality}. Dowe, P. and Salmon, M. H. (eds.). New York: Oxford University Press.

\bibitem[31]{Thrane} Thrane, E. Lasky, P. D., and Levin, Y. (2017)."Challenges for testing the no-hair theorem with current and planned gravitational-wave detectors." \emph{Physical Review D} 96, pp. 102004-1-102004-6.

\bibitem[32]{Walsh} Walsh J. L., Barth A. J., Ho Luis C. and Sarzi, M. (2013). “The M87 Black Hole Mass from Gas dynamical Models of Space Telescope Imaging Spectrograph Observations.” \emph{The Astrophysical Journal} 770, pp. 1-11.

\bibitem[33]{Weinstein} Weinstein, G. (2020). “Coincidence and reproducibility in the EHT black hole experiment.”
\emph{Studies in History and Philosophy of Science}, in press, pp. 1-16.

\bibitem[34]{Weinstein2} Weinstein, Galina (2021). "Demons in Black Hole Thermodynamics: Bekenstein and Hawking." \emph{ArXiv}:2021.11209.v1 [physics.hist-ph] 22 Feb.

\end{thebibliography}
\end{document}